\documentstyle[aps]{revtex}
\input{epsf.sty}
\newcommand{\be}{\begin{equation}}
\newcommand{\ee}{\end{equation}}
\newcommand{\omn}{\bar{\Omega}_n}
\newcommand{\sumn}{\sum_{n=-\infty}^{\infty}}
\newcommand{\sumk}{\sum_{k=0}^{\infty}}
\newcommand{\debye}{\Delta^{M}_{k,m}}
\newcommand{\sumup}{S_u^+ (\nu,t)}
\newcommand{\sumum}{S_u^- (\nu,t)}
\newcommand{\sumvp}{S_v^+ (\nu,t)}
\newcommand{\sumvm}{S_v^- (\nu,t)}
\newcommand{\mur}{\bar{\mu}}

\begin{document}

\title{Massless fermions in a bag at finite density and temperature}
\author{M.~De~Francia, H.~Falomir \\
Departamento de F\'{\i}sica, Facultad de Ciencias Exactas, Universidad Nacional de La Plata, \\
C.C. 67, 1900 La Plata, Argentina \\ ~~~~~~~~~~~~~~~ \\
M.~Loewe \\
Facultad de F\'{\i}sica, Pontificia Universidad Cat\'olica de Chile, \\
Casilla 306, Santiago 22, Chile\\
~~~~~~~~~~~}

\date{\today}

\maketitle

\vspace{-3in}
\hfill\vbox{
	\hbox{LA PLATA-TH 96/12}
	\hbox{hep-th/9608192}
}
\vspace{2.5in}

\begin{abstract}

We introduce the chemical potential in a system of massless fermions in a
bag by impossing boundary conditions in the Euclidean time direction. We
express the fermionic mean number in terms of a functional trace involving the
Green's function of the boundary value problem, which we study
analytically. Numerical evaluations are made, and an application to a
simple hadron model is discussed.

\bigskip 

\noindent
PACS number(s): 03.65.Db, 11.10.Wx, 12.39.Ba, 12.38.Mh

\end{abstract}

\section{Introduction}
\label{sec-introduccion}

Functional determinants of elliptic differential operators have wide
application in describing one-loop effects in Quantum and Statistical
Physics. When a bounded manifold in the Euclidean space-time is
considered, boundary conditions have a relevant influence on the behavior
of such determinants. 

In previous papers, we studied the relationship between
functional determinants of elliptic boundary value problems and the
corresponding Green's functions \cite{mochado,mitdef,fat}. This allowed for
the description of confined fields at finite temperature. 

In the framework of this functional techniques, we studied, in a
simplified scenario in two dimensions, the Gibbs free energy of confined
massless fermions \cite{chempot}. The chemical potential $\mu$ was
introduced through suitable boundary conditions impossed on the Dirac
operator in the Euclidean time direction. 

The present paper is an extension of that work: Here we consider a similar
calculation for a four dimensional M.I.T. bag model. In this effective
model of hadrons, a massless fermionic field is confined into a static
sphere of radius $R$ and subject to local spatial boundary conditions.
This problem has been considered previously \cite{mustafa} through an
asymptotic expansion of the energy density, for large values of the bag
radius. As we will show, the knowledge of the complete Green's function
for this boundary value problem allows us to describe the mean fermionic
number and the Gibbs free energy even at small radius, thus improving previous
approaches. 

The behavior of hadronic matter at finite temperature and density  has 
received considerable attention \cite{shuryak2} (see also \cite{tserruya} and references therein) during the last years. 
The main motivation behind this effort is an attempt to 
understand properly the generally accepted possibility of a 
deconfining phase transition from hadronic matter to the 
quark-gluon plasma \cite{smilga}, with applications to relativistic
heavy ion collisions and to the early Universe.

Different techniques have been used in 
connection with this problem. Among others, the lattice 
approach \cite{karsch}, the use of effective potentials for 
composite operators \cite{Barducci:1992}, chiral perturbation 
theory \cite{leutwyler}, the QCD Sum Rule method 
(see for example \cite{Dominguez:1994}), and  models like the 
Skyrmion approach and mixed 
constructions of hybrid models \cite{eskola,rhoreport,deconf,loewe} can be mentioned.

\bigskip

In section~\ref{sec-calculo} we obtain the fermionic mean number in terms
of a trace involving the Green's function of the problem. This last has a
nonregular behavior, which we study analytically in order to extract
physical results. 

In section~\ref{sec-aplicaciones} the difference between Gibbs and
Helmholtz ($\mu=0$ case) free energy is obtained. Complementing this
result with the Helmholtz energy previously evaluated in \cite{mitdef}, we
discuss its application to a simple quark-gluon droplet model
\cite{mustafa,mardor}. 

\section{The Gibbs free energy and mean fermionic number}
\label{sec-calculo}

In this section we will be interested in the study of the Gibbs 
free energy of a massless fermionic field confined inside a static 
sphere of radius $R$ and obeying local boundary conditions 
corresponding to the M.~I.~T. bag model. This amounts to study the 
functional determinant of the Dirac operator under such boundary 
conditions, which is related to the Grand canonical partition function 
according to 
\be
\Xi (T,R,\mu )=e^{-\beta G(T,R,\mu )} 
\sim Det\left( D(\beta ,R)-i\mu \gamma ^0\right) _{bc}\, .
\ee
Here
\be
D(\beta, R) = \frac i\beta \gamma ^0\partial _t+\frac iR  
\vec{\gamma}\cdot\vec{\nabla} ,\    
{\rm for\  }0\leq t,|\vec{x}|\leq 1,
\ee
and $"bc"$ means that the operator is defined on a space of functions 
satisfying the local boundary conditions 
\be
B \psi(t,\vec{x}) = \left({\bf 1}_4 + i \not \! n \right) \psi(t,\vec{x})=0 
\, ,
{\rm for \ } |\vec{x}|=1,
\ee
at the spatial edge and antiperiodic boundary conditions in the Euclidean
time direction,
\be
\psi(1,\vec{x})=-\psi(0,\vec{x})\, 
\ee
as corresponds to the fermionic character of the field.

In the above expression, $\mu$ stands for the chemical potential and 
$\beta$ denotes the inverse of the temperature. In our conventions, the 
Euclidean gamma matrices are

\be
\gamma_0 = i \rho^3 \otimes {\bf 1}_2 \, , 
\vec{\gamma} = i \rho^2 \otimes \vec{\sigma} \, ,
\gamma_5 = \rho^1 \otimes {\bf 1}_2 \, ,
\ee
where $\rho^k, \sigma^k, k=1,2,3,$ denote the Pauli matrices.

Following \cite{chempot}, we can relate $G(T,R,\mu )$ to the Green's 
function of this boundary value problem,
\begin{equation}  
\label{Dmu}   
\begin{array}{c}  
-\beta   
\frac{\partial G}{\partial \mu }(\beta ,R,\mu )=Tr\{\frac \partial {\partial  
\mu }\ln \left( D(\beta ,R)-i\mu \gamma ^0\right) _{bc}\} \\  \\   
=Tr\{-i\gamma ^0K_{bc}(t,x;t^{\prime },x^{\prime })\},   
\end{array} 
\label{numero} 
\end{equation}  
where $K_{bc}(t,x;t^{\prime },x^{\prime })$  satisfies    
\begin{equation}  
\label{Green}   
\begin{array}{c}  
\left( D(\beta ,R)-i\mu \gamma ^0\right) K_{bc}(t,\vec{x};
t^{\prime },\vec{x^{\prime}})=
\delta (\vec{x}-\vec{x^{\prime }})\delta (t-t^{\prime }) \\    
\\   
BK_{bc}(t,\vec{x};t^{\prime },\vec{x^{\prime }})=0,   
{\  \rm  for \  }|\vec{x}|=1, \\  \\   
K_{bc}(1,\vec{x};t^{\prime },\vec{x^{\prime }})=
-K_{bc}(0,\vec{x};t^{\prime },\vec{x^{\prime }}).   
\end{array}  
\end{equation} 
In fact,  (\ref{numero}) gives us the particle mean number 
$N(\beta, \mu)= - {{\partial G}\over{\partial \mu}}$.

As discussed in \cite{chempot}, the chemical potential $\mu$ 
can also be included in the boundary condition for the Euclidean 
time direction, through the invertible transformation 

\begin{equation}  
\begin{array}{c}  
D(\beta ,R)-i\mu \gamma ^0=e^{\mu \beta t}D(\beta ,R)\ e^{-\mu \beta t}, \\  
\\   
K_{bc}(t,\vec{x};t^{\prime },\vec{x^{\prime }})=e^{\mu \beta t}k(t,\vec{x};
t^{\prime  
},\vec{x^{\prime }})\ e^{-\mu \beta t\prime },   
\end{array}  
\end{equation}  
where $k(t,\vec{x};t^{\prime },\vec{x^{\prime }})$ is the Green's 
function of the modified problem,   
\begin{equation}  
\begin{array}{c}  
D(\beta ,R)\ k(t,\vec{x};t^{\prime },\vec{x^{\prime }})=\delta 
(\vec{x}-\vec{x^{\prime }})\delta  
(t-t^{\prime }) \\    
\\   
Be^{\mu \beta t}k(t,\vec{x};t^{\prime },\vec{x^{\prime }})=0,   
{\  \rm  for \   }|\vec{x}|=1, \\  \\   
k(1,\vec{x};t^{\prime },\vec{x^{\prime }})+e^{-\mu  
\beta }k(0,\vec{x};t^{\prime },\vec{x^{\prime }})=0.   
\end{array} 
\label{greencom} 
\end{equation}  

This function has the development   

\begin{equation}  
\label{desk}k(t,x;t^{\prime },x^{\prime })=R \sum\limits_{n=-\infty }^\infty  
k_n(x,x^{\prime })\ e^{-i\Omega _n\beta (t-t^{\prime })},   
\end{equation}  
with the frequencies given by   
\begin{equation}  
\label{Omega}   
\begin{array}{c}  
\Omega _n=\omega _n-i\mu , \\    
\\   
\omega _n={\displaystyle {{(2n+1)\pi }\over\beta }},{\  \rm  for \  } 
n\in {\bf Z,}   
\end{array}  
\end{equation}
and $k_n(x;x^{\prime })$ satisfying
\be
\left[\Omega_n R i \rho^3 \otimes {\bf 1}_2 -
\rho^2 \otimes \vec{\sigma}\cdot \vec{\nabla} \right] 
k_n(x;x^{\prime}) = \delta^{(3)} (x-x^{\prime}) \rho^0 \otimes {\bf 1}_2,
\label{ec-dif}
\ee
\be
B k_n(\vec{x};\vec{x^{\prime }})=0. 
\ee
It is convenient to define dimensionless parameters $\bar{\Omega}_n = R 
\Omega_n$, $\bar{\mu} = R \mu$ and $z=R/\beta$.

\bigskip

The solution of (\ref{greencom}) can be expressed as the sum of a solution
of the inhomogeneous equation, $k_{0}(t,x;t^{\prime },x^{\prime })$, with
a vanishing behavior at infinity, and a solution of the homogeneous
equation, $\tilde{k}(t,x;t^{\prime },x^{\prime })$, chosen in such a way
that the complete Green's function $k(t,x;t^{\prime },x^{\prime })$ obeys
the boundary condition in (\ref{greencom}). This splitting of the Green's
function induces a similar separation for the mean fermionic number.  In
subsections \ref{subsec-libre} and \ref{subsec-contorno}, we will
considered both pieces separately. 

\subsection{Fermionic mean number for the boundaryless case}
\label{subsec-libre}

In this case the mean fermionic number is given by

\be
N_0 = - \left. \frac{\partial G}{\partial \mu} \right|_{0} =
\frac1{\beta} {\rm Tr} \left[ -i \gamma^0 e^{\mu \beta t} k_0 e^{- \mu \beta 
t'} \right]\, ,
\label{numero0}
\ee
where $k_0$ corresponds to the Green's function of the modified 
problem (\ref{greencom}), satisfying the temporal boundary condition and
vanishing at spatial infinity,

\[  
k_0(t,x;t^{\prime },x^{\prime })=R \sum\limits_{n=-\infty }^\infty  
k_0^{(n)}(x,x^{\prime })\ e^{-i\Omega _n\beta (t-t^{\prime })},   
\]
\be
k_0^{(n)} (\vec{x},\vec{x'}) = 
- \frac1{4 \pi} 
\frac{e^{-S_n \bar{\Omega}_n \left| \vec{x} - \vec{x'} \right| }}{\left| 
\vec{x} - \vec{x'} \right|} 
\left[ \bar{\Omega}_n \gamma^0 +
\left( \frac1{\left| \vec{x} - \vec{x'} \right|} - S_n \bar{\Omega}_n \right)
\frac{i \vec{\gamma} \cdot \left( \vec{x} - \vec{x'} \right)}{\left| \vec{x} 
- \vec{x'} \right|}
\right]\, .
\ee
In the previous expression, $S_n = {\rm sign} (n+1/2)$.
 
In order to get $N_0$ we have to compute the trace in eq. (\ref{numero0}). 
Due to the singular behavior of $k_0$ at the diagonal, we can take $t'=t$, 
but we should keep $\vec{x'} \neq \vec{x}$ up to the end of the 
calculation. In this way,

\[
N_0 =\frac1{\beta} \int_0^1 dt \int_0^1 r^2 dr
\int_{\Omega} d\Omega \,
\left\{ {\rm tr} \left[ -i \gamma^0 R \sum_{n=-\infty}^{\infty} 
\left( - \frac1{4 \pi} \right) 
\frac{e^{-S_n \bar{\Omega}_n \left| \vec{x} - \vec{x'} \right|}}{\left| 
\vec{x} - \vec{x'} \right|} \times \right. \right.
\]
\be
\left. \left.  \left\{  {\bar{\Omega}_n \gamma^0} +
\left( \frac1{\left| \vec{x} - \vec{x'} \right|} - S_n \bar{\Omega}_n \right)
\frac{i \vec{\gamma} \cdot \left( \vec{x} - \vec{x'} \right)}{\left| 
\vec{x} - \vec{x'} \right|}  \right\}  \right] \right\}_{\vec{x'} \rightarrow 
\vec{x}}\, .
\label{traza}
\ee
The second term in the integrand of (\ref{traza}) vanishes because of the 
properties of Dirac matrices. For the first one, with 
$\epsilon = \left|\vec{x} - \vec{x'} \right|$, we get

\[
N_0 = \lim_{\epsilon \rightarrow 0^+} 
- \frac43 \frac{i R}{\epsilon}
\sum_{n=-\infty}^{\infty} \bar{\Omega}_n e^{- \epsilon S_n
\bar{\Omega}_n} =
\]
\be
\left. - \frac8{3 \beta} \frac{R}{\epsilon} \frac{\partial}{\partial \epsilon}
\Im \sum_{n=0}^{\infty} e^{- \epsilon \bar{\Omega}_n} \right|_{\epsilon 
\rightarrow 0^+}
\ee
So,

\be
N_0={{4\pi}\over{9}} \left( \bar{\mu} z^2 + {{\bar{\mu}^3}\over{\pi^2}} \right),
\label{ene0}
\ee
in agreement with the value for a free fermionic gas, as can be seen, for instance, in Ref.~\cite{gases}. 

\subsection{Influence of the boundary conditions on the fermionic mean number}
\label{subsec-contorno}

For the solution of the homogeneous equation required to adjust the boundary 
conditions for the full Green's function, we propose the development
\be
\tilde{k}^{(n)}( \vec{x} , \vec{x'} )=\sum_{k=0}^{3}
\rho^k \otimes a^k ( \vec{x} , \vec{x'} )\, ,
\label{ktilde}
\ee
and write $k_0^{(n)}( \vec{x} , \vec{x'} )$ as
\be
k_0^{(n)}( \vec{x} , \vec{x'} )=\sum_{k=0}^{3}
\rho^k \otimes A^k ( \vec{x} , \vec{x'} )\, .
\label{kacero}
\ee

Concerning the $A^k$ coefficients, $A^0$ and $A^1$ vanish, 
since they are the solutions of an everywhere regular homogeneous 
problem. For the other coefficients we have

\be
A^2( \vec{x} , \vec{x'} )=-{{\vec{\sigma}\cdot\vec{\nabla}}\over{i 
\bar{\Omega}_n}} A^3( \vec{x} , \vec{x'} ) ,
\ee

\be
\left[\nabla^2 + \left( i \bar{\Omega}_n \right)^2 \right] A^3( 
\vec{x} , \vec{x'} ) =  i \bar{\Omega}_n \delta^{(3)}( \vec{x} , 
\vec{x'}) {\bf 1}_4 .
\ee

Following \cite{hanson}, we get $A^3$ as 
an expansion in terms of spinorial spherical harmonics,

\be 
A^3( \vec{x} , \vec{x'})=\sum_{j,l,l',m}A^3_{j,l,l',m}(r,r')\phi_{j,l,m}(\Omega)
{\phi^\dagger}_{j,l',m}(\Omega'),
\ee
where
\be
A^3_{j,l,l',m}(r,r')=\delta_{l,l'} i S_n \bar{\Omega}_n^2 j_l(i S_n \bar{
\Omega}_n r_<) h_{l'}^{(1)}(i S_n \bar{\Omega}_n r_>) \, .
\label{A3}
\ee
In eq. (\ref{A3}), $j_l(x)$ and $h_l^{(1)}(x)$ are the 
spherical Bessel and Hankel functions, respectively, and 
$r_<={\rm Min}(r,r')$ ($r_>={\rm Max}(r,r')$).

For the coefficients $a^k$ in eq. (\ref{ktilde}), we get

\be
\begin{array}{cc}
a^2=-{\displaystyle{{{\vec{\sigma}\cdot\vec{\nabla}}\over{i \bar{
\Omega}_n}}}}a^3, &
a^1=-{\displaystyle{{{\vec{\sigma}\cdot\vec{\nabla}}\over{i \bar{
\Omega}_n}}}}a^0,\\
 & \\
\left[\nabla^2 + \left( i \bar{\Omega}_n \right)^2 \right]a^3=0, &
\left[\nabla^2 + \left( i \bar{\Omega}_n \right)^2 \right]a^0=0,
\end{array}
\ee
while the boundary conditions at $r=R$ imply

\be
\begin{array}{c}
\vec{\sigma}\cdot\hat{r} a^0 -a^2 = A^2,\\
\vec{\sigma}\cdot\hat{r} a^1 -i a^3 = i A^3.
\end{array}
\ee

Using the following Ansatz,

\be 
a^k(\vec{x},\vec{x'})=\sum_{j,l,l',m} C_{j,l,l',m}^{k} j_l (i S_n 
\bar{\Omega}_n r) j_l' (i S_n \bar{\Omega}_n r')\phi_{j,l,m}(\Omega)
{\phi^\dagger}_{j,l',m}(\Omega'),
\ee
we finally get

\[
k_n (\vec{x},\vec{x}') = (i \omn )^2 \sum_{j,l,l',m}
\]
\[
\left[
\left(
\rho^0 \delta_{l,l'} i^{\bar{l} - l} d_j+ 
\rho^1 \delta_{\bar{l},l'} S_n i^{\bar{l} - l} d_j -
\rho^2 \delta_{\bar{l},l'} c_j +
\rho^3 i S_n \delta_{l,l'} c_j
\right) j_l (i S_n \omn r ) j_{l'} (i S_n \omn r') 
\right.
\]
\be
+ \left. 
\left(
\rho^2 \delta_{\bar{l},l'} -
\rho^3 i S_n \delta_{l,l'}
\right) j_l (i S_n \omn r_< ) h_{l'}^{(1)} (i S_n \omn r_>)
\right] \otimes
\phi_{j,l,m}(\Omega)
{\phi^\dagger}_{j,l',m}(\Omega') \! \!
\, ,
\ee
where $c_j$ and $d_j$ are given by

\[
c_j={{j_{j+1/2}(i S_n \bar{\Omega}_n) h_{j+1/2}^{(1)}(i S_n 
\bar{\Omega}_n)- j_{j-1/2}(i S_n \bar{\Omega}_n) h_{j-1/2}^{(1)}(i 
S_n \bar{\Omega}_n)}
\over{j_{j+1/2}^2(i S_n \bar{\Omega}_n) -j_{j-1/2}^2(i S_n \bar{\Omega}_n)}},
\]

\be
d_j= {{-i/\bar{\Omega}_n^2}\over{{j_{j+1/2}^2(i S_n \bar{\Omega}_n) -
j_{j-1/2}^2(i S_n \bar{\Omega}_n) }}}.
\ee

The correction to the fermionic mean number due to the boundary conditions is
\be
\tilde{N} = 
\frac1{\beta} {\rm Tr} \left[ -i \gamma^0 e^{\mu \beta t} \tilde{k} 
e^{- \mu \beta t'} \right].
\label{numerotilde}
\ee
Once again, in order to get a well defined result, 
we must proceed carefully when treating the kernel of this 
operator at the diagonal. We can take $t'=t$ and $\Omega'=\Omega$ 
safely, while keeping $r'=r(1-\epsilon)$, with $\epsilon >0$. We thus get

\[
\tilde{N}=2 z \lim_{\epsilon \rightarrow 0^+} \sum_{n=-\infty}^{\infty} i 
S_n (i \bar{\Omega}_n)^2 \sum_{j,m} c_j \int_0 ^1 r^2 \, dr \times
\]
\be
\left[
j_{j+1/2}^2(i S_n \bar{\Omega}_n r)j_{j+1/2}^2(i S_n \bar{\Omega}_n r')
+ j_{j-1/2}^2(i S_n \bar{\Omega}_n r)j_{j-1/2}^2(i S_n \bar{\Omega}_n r') 
\right]_{r'=r(1-\epsilon)}.
\ee

After a straightforward calculation of the radial integral, 
we obtain the following expresion 
\[
\tilde{N} = \lim_{\epsilon \rightarrow 0^+}
\sum_{n=-\infty}^{\infty} N_{n,k} (z, \bar{\mu}; \epsilon) 
\]
\[= -2 i z \lim_{\epsilon \rightarrow 
0^+} \sumn \sumk \frac{(1-\epsilon)^{-3/2}}{\epsilon}
\frac{1}{\bar{\Omega}_n}
\left( 2 \nu +1 \right)
\frac{d_{\nu,n}^{\epsilon} - d_{\nu,n}
+ \epsilon \left( d_{\nu,n} - \nu \right)}{\left( d_{\nu,n} - \nu \right)^2 
+ \bar{\Omega}_n^2} \times
\]
\[
\frac{I_\nu \left(S_n \omn \left( 1 - \epsilon \right) \right)}
{I_\nu \left(S_n \omn \right)} 
\left[ 
\omn^2 I_{\nu}'( S_n \omn )
K_{\nu}' ( S_n \omn ) - S_n \omn \nu \left( I_\nu (S_n \omn) 
K_\nu (S_n \omn ) \right)' \right.
\]
\be
\left.
+
\left( \nu^2 + \omn^2 \right) I_\nu \left( S_n \omn \right) 
K_\nu \left( S_n \omn \right) \right]\, ,
\label{predebye}
\ee
where
\be
d_{\nu,n}^{\epsilon} = \left. S_n \omn (1 - \epsilon ) 
\frac{d}{dx} \log I_\nu \left( x \right) \right|_{x= S_n 
\omn (1 - \epsilon)}\, ,
\qquad
\nu = k+\frac12 \, ,
\ee
and $\left. d_{\nu,n} = d_{\nu,n}^{\epsilon} \right|_{\epsilon=0}$.

The double series (\ref{predebye}) is not absolutely 
convergent for $\epsilon = 0$, so we must keep $\epsilon > 0$ in the 
general term. A way to isolate such singular behavior consists 
in the substraction of the first terms in the 
asymptotic (Debye) expansion of $N_{k,n}$, $\Delta^M_{k,n}$ (see, for example, 
\cite{mitdef}). The difference 
\be
\left| N_{k,n} (z, \bar{\mu};\epsilon ) - \Delta^{M}_{k,m} (z, \bar{\mu};
\epsilon)\right| \asymp {\cal O} \left( \frac{1}{\rho^{M+1}} \right) \, ,
\ee
where $\rho^2 = \omn^2 + \nu^2$ can be considered as a ``radial" 
variable in the double series.   

So, we can write
\[
\tilde{N} = \tilde{N}_{1} + \tilde{N}_{2} =
\]
\be
\sumn \sumk \left( N_{k,n} (z, \bar{\mu};\epsilon ) - \Delta^{M}_{k,m} (z, 
\bar{\mu};\epsilon) \right)_{\epsilon = 0} + \lim_{\epsilon \rightarrow 0}
\sumn \sumk \Delta^{M}_{k,n} (z,\bar{\mu};\epsilon) \, .
\label{debye}
\ee
In the first term of (\ref{debye}), for $M \geq 3$, the limit 
$\epsilon \rightarrow 0$ can be taken for the general term, 
inside the (absolutely and uniformly convergent) double sum. 
The nonregular behavior of the series is isolated in the second term of 
(\ref{debye}).

Our calculation strategy consists in the numerical evaluation of the 
first term of (\ref{debye}), and the analytical study of the 
second one. Notice that we may take $M>3$ in order to improve 
the numerical convergence of the first term.

\subsubsection*{Asymptotic (Debye) expansions}

Introducing the Debye expansion for the modified Bessel 
functions \cite{abram}, we can write
\be
\debye = \omn D(\nu,t,\epsilon) E(\nu,t,\epsilon) \, ,
\ee
with
\[
D(\nu,t,\epsilon ) =\left\{
(- 2 i z) (1-\epsilon)^{-3/2} \quad \frac{2 \nu + 1}{2 \nu} \quad
\frac{d_{\nu,n}^\epsilon - d_{\nu,n} + \epsilon \left( d_{\nu,n} - 
\nu \right)}{\epsilon} \times \right.
\]
\[
\frac1{d_{\nu,n}^2 - 2 \nu d_{\nu,n} + \nu^2 / t^2 } 
\quad f^{-1/4} (\epsilon,t) \quad
\frac{S_u^+ \left( \nu,t f^(-1/2) (\epsilon,t) 
\right)}{S_u^+ \left( \nu,t \right)} \times
\]
\be
\frac{t}{1-t^2} \left[ \sumup \sumum - \sumvp \sumvm + \right. 
\left. \left.
\left.
t \left( \sumup \sumvm - \sumum \sumvp \right) \right]
\begin{array}{c}
\\ \\
\end{array} \right\} \right|_M
\ee
where $\left. \right|_M$ stands for a consistent expansion up 
to the order $\rho^{-M}$, and
\be
E(\nu,t,\epsilon) =
\exp \left( \frac{\nu}{t} \left( f^{1/2} (\epsilon,t) -1 \right) +
\nu \log \left( \left( 1-\epsilon \right) \frac{t+1}{t + f^{1/2} 
(\epsilon,t)} \right) \right).
\ee
In the expressions above, 
\[
t=\frac{\nu}{\rho} \, , \qquad f(\epsilon , t) = 1 - \epsilon \left(2 - 
\epsilon \right) \left( 1 - t^2 \right) \, ,
\]
\begin{eqnarray}
\sumup &=& 1 + \sum_{k=1}^{\infty} \frac{u_{k} (t)}{\nu^k} \, , \\
\sumum &=& 1 + \sum_{k=1}^{\infty} (-1)^k \frac{u_{k} (t)}{\nu^k} \, , \\
\sumvp &=& 1 + \sum_{k=1}^{\infty} \frac{v_{k} (t)}{\nu^k} \, , \\
\sumvm &=& 1 + \sum_{k=1}^{\infty} (-1)^k \frac{v_{k} (t)}{\nu^k} \, ,
\label{sumas}
\end{eqnarray}
with $u_{k} (t)$ and $v_{k} (t)$ being the polynomials 
appearing in the Debye expansion of Bessel functions, 
as defined as in Ref. \cite{abram}.

It is easy to see that
\be
E(\nu,t,\epsilon) = \exp \left( - \epsilon \frac{\nu}{t} \right)
\left( 1 + O(\epsilon^2) \right) \,.
\label{factor_e}
\ee
So, the corrections to the fermionic mean number due to the 
boundary conditions (\ref{debye}) can be written as
\be
\tilde{N}_1 =
\sumn \sumk \left[ N_{n,k} (z,\bar{\mu};\epsilon =0) -
\omn D(\nu,t,\epsilon =0) \right]
\label{eneuno} \, ,
\ee
\be
\tilde{N}_2 =
\left[ \sumn \sumk \omn D(\nu,t,\epsilon) E(\nu,t,\epsilon) 
\right]_{\epsilon \rightarrow 0^+} \, .
\label{factor_d}
\ee

The first term, $\tilde{N}_1$, will be numerically 
evaluated from the expression
\[
N_{n,k} (z,\bar{\mu},\epsilon=0 ) =
\]
\[
(-2 i z) \omn \left( 2 \nu +1 \right) \frac{ d_{\nu,n}^2 - 
\nu^2 / t^2 + d_{\nu,n} - \nu}{d_{\nu,n}^2 - 2 \nu d_{\nu,n} + \nu^2 / t^2}  
\left[ I_{\nu}'( S_n \omn )K_{\nu}' ( S_n \omn )\right.
\]
\be
\left.
 - \frac{\nu}{S_n \omn} \nu \left( I_\nu (S_n \omn) 
K_\nu (S_n \omn ) \right)' +
\frac{\nu^2 + \omn^2}{\omn^2} I_\nu \left( S_n \omn \right) 
K_\nu \left( S_n \omn \right) \right]
\label{general}
\ee
and from the asymptotic expansion $D$, at $\epsilon =0$. For 
computational convenience we take $M=6$; the 
corresponding expression for $D$ is
\[
D(\nu,t,\epsilon=0)= -2 i z \left[
{{-{t^5}}\over {2\,{{\nu }^2}}} + 
  {1\over {{{\nu }^3}}}\left(
{{{-3\,{t^5}}\over 4} - {{{t^6}}\over 2} + {{5\,{t^7}}\over 4} + 
      {{3\,{t^8}}\over 4}} \right)  \right.
\]
\[
+{1\over {{{\nu }^4}}}
\left( {{{-3\,{t^5}}\over 8} - {{3\,{t^6}}\over 4} + 
      {{11\,{t^7}}\over {16}} + {{27\,{t^8}}\over 8} + 
      {{7\,{t^9}}\over 2} - 3\,{t^{10}} - {{71\,{t^{11}}}\over {16}}} \right)  
\]
\[
+{1 \over {{{\nu }^5}}} \left(
{{-{t^5}}\over {16}} - 
      {{3\,{t^6}}\over 8} - {{27\,{t^7}}\over {32}} + 
      {{9\,{t^8}}\over 4} + {{483\,{t^9}}\over {32}} + 
      {{25\,{t^{10}}}\over 8} 
\right.
\left.
- {{1211\,{t^{11}}}\over {32}} - 
      {{285\,{t^{12}}}\over {16}} + {{781\,{t^{13}}}\over {32}} + 
      {{213\,{t^{14}}}\over {16}}
\right) 
\]
\[ 
+  {1\over {{{\nu }^6}}} \left(
{{-{t^6}}\over {16}} - {{47\,{t^7}}\over {64}} - 
      {{3\,{t^8}}\over 4} + {{2909\,{t^9}}\over {256}} + 
      {{473\,{t^{10}}}\over {16}} - {{531\,{t^{11}}}\over {64}} 
\right.
\]
\be
\left.
\left.
- 
      {{4255\,{t^{12}}}\over {32}} - {{14795\,{t^{13}}}\over {128}} + 
      {{6297\,{t^{14}}}\over {32}} + {{481\,{t^{15}}}\over 2} - 
      {{1491\,{t^{16}}}\over {16}} - {{32799\,{t^{17}}}\over {256}}
     \right)\right] \, .
\label{desdeb}
\ee

\bigskip 

The second term, $\tilde{N}_2$, will be studied 
analytically in the following. Taking into acount (\ref{factor_e}), we have
\be
\tilde{N}_2 =\left[
\sumn \sumk \omn
D (\nu,t,\epsilon = 0) e^{-\epsilon \nu /t} \left( 1 + O(\epsilon) \right)
\right]_{\epsilon \rightarrow 0^+} \, ,
\label{enedos}
\ee

Now, employing (\ref{desdeb}), Equation (\ref{enedos}) 
can be naturally expressed in terms of the double series
\be
s(p,q; \epsilon ) =
\sumn \sumk
(- 2 z) i\omn e^{-\epsilon \rho } \rho^{-p} \nu^{q}\, ,
\label{sds}
\ee
(where, for convenience, we omit the dependence on $\bar{\mu}$ and $z$)  
obtaining
\[
\tilde{N}_2= \lim_{\epsilon \rightarrow 0^+}
\left[ \left\{
{{-s(5,0;\epsilon)}\over {16}} - 
  {{3\,s(5,1;\epsilon)}\over 8} - 
  {{3\,s(5,2;\epsilon)}\over 4} - 
  {{s(5,3;\epsilon)}\over 2} 
\right. \right.
\]
\[
\left.
- 
  {{s(6,0;\epsilon)}\over {16}} - 
  {{3\,s(6,1;\epsilon)}\over 8} - 
  {{3\,s(6,2;\epsilon)}\over 4} - 
  {{s(6,3;\epsilon)}\over 2} 
\right.
\]
\[
\left.- 
  {{47\,s(7,1;\epsilon)}\over {64}} - 
  {{27\,s(7,2;\epsilon)}\over {32}} + 
  {{11\,s(7,3;\epsilon)}\over {16}} + 
  {{5\,s(7,4;\epsilon)}\over 4} 
\right.
\]
\[
\left.
- 
  {{3\,s(8,2;\epsilon)}\over 4} + 
  {{9\,s(8,3;\epsilon)}\over 4} + 
  {{27\,s(8,4;\epsilon)}\over 8} + 
  {{3\,s(8,5;\epsilon)}\over 4} 
\right.
\]
\[
\left.
+ 
  {{2909\,s(9,3;\epsilon)}\over {256}} + 
  {{483\,s(9,4;\epsilon)}\over {32}} + 
  {{7\,s(9,5;\epsilon)}\over 2} + 
  {{473\,s(10,4;\epsilon)}\over {16}} 
\right.
\]
\[
\left.
+ 
  {{25\,s(10,5;\epsilon)}\over 8} - 
  3\,s(10,6;\epsilon) - 
  {{531\,s(11,5;\epsilon)}\over {64}} - 
  {{1211\,s(11,6;\epsilon)}\over {32}} 
\right.
\]
\[
\left.
- 
  {{71\,s(11,7;\epsilon)}\over {16}} - 
  {{4255\,s(12,6;\epsilon)}\over {32}} - 
  {{285\,s(12,7;\epsilon)}\over {16}} - 
  {{14795\,s(13,7;\epsilon)}\over {128}} 
\right.
\]
\[
\left.
+ 
  {{781\,s(13,8;\epsilon)}\over {32}} + 
  {{6297\,s(14,8;\epsilon)}\over {32}} + 
  {{213\,s(14,9;\epsilon)}\over {16}} + 
  {{481\,s(15,9;\epsilon)}\over 2} 
\right.
\]
\be
\left. \left.
- 
  {{1491\,s(16,10;\epsilon)}\over {16}} - 
  {{32799\,s(17,11;\epsilon)}\over {256}}
\right\}
(1 + {\cal O} (\epsilon )) \right]\, .
\label{enedosdes}
\ee

In the following section we will prove that 
$s(p,q;\epsilon)$ has a finite $\epsilon 
\rightarrow 0^+$-limit, which allows for the evaluation of (\ref{enedosdes}). 

\subsubsection*{Evaluation of the double series $s(p,q;\epsilon)$}

In this section we study the double sums defined in 
equation (\ref{sds}), which can be written as
\be
s(p,q; \epsilon ) =
\sumk \nu^q {\cal S}_p (\nu, \epsilon)\, ,
\label{seriedoble}
\ee
where
\be
{\cal S}_p (\nu, \epsilon) =
(- 2 z) \sumn
 \frac{i\omn e^{-\epsilon \sqrt{\nu^2 + \omn^2}}}{\left(\nu^2 + 
\omn^2 \right)^{p/2}}\, . 
\label{serie_n}
\ee

It is easily seen that the ${\cal S}_p (\nu, \epsilon)$ are 
odd-functions of $\mur$, and satisfy the recursion relations
\be
{\cal S}_{p-1} (\nu, \epsilon)=
- \frac{\partial}{\partial \epsilon} {\cal S}_p (\nu, \epsilon)\, ,
\label{rec_ep}
\ee
\be
{\cal S}_p (\nu, \epsilon)=
- \frac{1}{m-2} \left[ \epsilon {\cal S}_{p-1} (\nu, \epsilon) + 
\frac1\nu \frac{\partial {\cal S}_{p-2} (\nu, \epsilon)}{\partial \nu} 
\right], \qquad
{\rm for} \quad m \neq 2 \, ,
\label{rec_nu}
\ee
which allow the computation of any ${\cal S}_p (\nu, \epsilon)$ 
from the knowledge of ${\cal S}_2 (\nu, \epsilon)$. The last series 
will be determined in what follows.

Making use of the Poisson formula \cite{abram}, 
${\cal S}_2 (\nu, \epsilon)$ can be transformed into the series

\be
{\cal S}_2 (\nu, \epsilon) = \sum_{\ell=-\infty}^{\infty} C_\ell \, ,
\ee
where the coefficients are given by

\be
C_\ell = \frac{(-1)^\ell}{\pi} 2\Im \int_{0}^{\infty} \, dx \, 
\frac{x - i \bar{\mu}}{(x - i \bar{\mu})^2 + \nu^2}
e^{i \ell x/z} e^{- \epsilon \sqrt{(x - i \bar{\mu})^2 + \nu^2}}\, .
\ee

It is convenient to treat the terms with $\ell =0 , \ell >0$ and 
$\ell < 0$ separately. The corresponding integrals can be 
exactly computed in each case, by adequately extending the 
integration path on the complex $x$-plane. For $\bar{\mu}>0$, one 
gets

\be
C_0 = H(\bar{\mu}-\nu)\left[ 1-{2\over\pi} \int_\nu^{\bar{\mu}} \, dx 
{x \over {x^2 - \nu^2}} \sin\left(\epsilon\sqrt{x^2-\nu^2}\right)
\right] \, ,
\label{C0}
\ee

\bigskip

\[
C_{\ell>0} = (-1)^\ell \left\{ e^{-\ell (\bar{\mu}+\nu)/z} - {2\over \pi}
\int_\nu^\infty \, dx {x\over{x^2-\nu^2}} e^{-\ell (\bar{\mu}+x)/z}
\sin(\epsilon \sqrt{x^2-\nu^2}) \, + \right.
\]
\be \left.
 H(\bar{\mu}-\nu)\left[e^{-\ell (\bar{\mu}-\nu)/z}- {2\over \pi}
\int_\nu^{\bar{\mu}} \, dx {x\over{x^2-\nu^2}} e^{-\ell (\bar{\mu}-x)/z}
\sin(\epsilon \sqrt{x^2-\nu^2})\right] \right\}\, , 
\label{Cpos}
\ee

\bigskip

\[
C_{\ell<0} = (-1)^\ell \left\{ {2\over \pi}H(\bar{\mu}-\nu)
\int_{\bar{\mu}}^\infty \, dx {x\over{x^2-\nu^2}} e^{-\ell (\bar{\mu}-x)/z}
\sin(\epsilon \sqrt{x^2-\nu^2}) \, + \right.
\]
\be \left.
 H(\nu-\bar{\mu})\left[-e^{-\ell (\bar{\mu}-\nu)/z}+ {2\over \pi}
\int_\nu^\infty \, dx {x\over{x^2-\nu^2}} e^{-\ell (\bar{\mu}-x)/z}
\sin(\epsilon \sqrt{x^2-\nu^2})\right] \right\}\, , 
\label{Cneg}
\ee
where $H(x)$ stands for the Heaviside step function.

Finally, the sum over the index $\ell$ can be perfomed to get 

\[
{\cal S}_2 (\nu, \epsilon) = {1\over{1+e^{-(\bar{\mu}-\nu)/z}}} - 
{1\over{1+e^{(\bar{\mu}+\nu)/z}}}\, - 
\]

\be
{2\over \pi}
\int_0^\infty \, {du\over u} \left[  {1\over{1+e^{-\bar{\mu}/z}
e^{{\nu \over z}\sqrt{u^2 +1}}}} - 
{1\over{1+e^{+\bar{\mu}/z}e^{{\nu \over z}\sqrt{u^2 +1}}}}\right] 
\sin(\epsilon\nu u)\, .
\label{s2}
\ee   

Now, the expressions for ${\cal S}_{p\neq 2} (\nu, \epsilon)$ 
can be inmediately obtained employing (\ref{rec_ep}) and (\ref{rec_nu}). 

\bigskip

All ${\cal S}_{p} (\nu, \epsilon)$ so defined are 
regular at $\epsilon =0$, and exponentially vanishing 
with $\nu$ (even at $\epsilon=0$). Such behavior guaranties 
the uniform convergence of the $k$-series in 
expression (\ref{seriedoble}), allowing to take the
$\epsilon\rightarrow 0$ limit of ${\cal S}_p (\nu, \epsilon)$ therein,
\be
s(p,q; \epsilon = 0) =
\sumk \nu^q {\cal S}_p (\nu, \epsilon = 0)\, .
\label{limite}
\ee
This limit also simplifies the recursion relations leading to

\be
{\cal S}_{2 \kappa} (\nu, 0) = 
\frac{1}{[2(\kappa -1)]!!} \left( - \frac1\nu 
\frac{\partial}{\partial \nu} \right)^{\kappa -1} {\cal S}_{2} (\nu, 0)
\label{s2k}
\ee
\be
{\cal S}_{2 \kappa + 1} (\nu, 0) = 
\frac{1}{[2 \kappa -1]!!} \left( - \frac1\nu 
\frac{\partial}{\partial \nu} \right)^{\kappa } {\cal S}_{1} (\nu, 0) \, ,
\label{s2kp1}
\ee
where
\be
{\cal S}_1 (\nu, 0) = {{2\nu}\over \pi}
\int_0^\infty \, du \left[  {1\over{1+e^{-\bar{\mu}/z}
e^{{\nu \over z}\sqrt{u^2 +1}}}} - 
{1\over{1+e^{+\bar{\mu}/z}e^{{\nu \over z}\sqrt{u^2 +1}}}}\right] \, ,
\label{ese1}
\ee
\be
{\cal S}_2 (\nu,0) = {1\over{1+e^{-(\bar{\mu}-\nu)/z}}} - 
{1\over{1+e^{(\bar{\mu}+\nu)/z}}} \, .
\label{ese2}
\ee

\bigskip

Equations (\ref{s2k}) to (\ref{ese2}) lead to 
exact expressions for ${\cal S}_{p} (\nu, 0)$, allowing 
for the numerical evaluation of $s(p,q; \epsilon = 0)$. 
We can also obtain analytically an 
expansion of $s(p,q; \epsilon = 0)$ for large values of $z=RT$.
In fact, for $p \ge 4$ it is straightforward to get
\be
s(p,q; \epsilon = 0) = (-2 z) \frac{1}{p-2}
\frac{\partial}{\partial \mur} 2 \Re \sigma_{p-2,q} \, ,
\ee
where
\be
\sigma_{p,q} = \sumk \nu^q \sum_{n=0}^{\infty}
\rho^{-p}
\ee

Making use of the integral representation
\[
\rho^{-p} =
\frac1{\Gamma \left( p/2 \right)}
\int_0^\infty \, d\tau \, \tau^{\frac{p}{2} -1}
e^{-\tau \rho^2}
\]
we obtain
\be
\sigma_{p,q} =
\frac1{\Gamma \left( p/2 \right)}
\int_0^\infty \, d\tau \, \tau^{\frac12 (p-q) -1}
\left( \sumk \left( \tau \nu^2 \right)^{q/2} e^{-\tau \nu^2} \right)
\left( \sum_{n=0}^{\infty} 
e^{-\tau \left(2 \pi z \right)^2 \left[ n + 
\left( \frac12 - i \frac{\mur}{2 \pi z} \right) \right]^2 } \right)\, .
\label{high}
\ee
In this expression, for large $z$ and for each $n$, 
high values of $\tau$ are strongly suppressed by the 
exponential factor $e^{-\tau \left(2 \pi z \right) 
\left( n +  \frac12 \right)}$. So, we can use 
the Euler-Mac Laurin expansion \cite{abram} to evaluate the factor
\be
\sumk
\left(\tau \nu^2 \right)^{q/2} e^{-\tau \nu^2}  =
\frac12 \frac1{\sqrt{\tau}} \Gamma \left( \frac{d+1}{2} \right)
+ \frac12 \sum_{\ell=1}^{\infty} \frac{(-1)^{\ell +1}}{(\ell + 1)!}
\left[ \frac1{2^\ell } - 2 B_{\ell +1} \right]
\tau^{\ell /2} \frac{d^\ell}{dx^\ell} \left. \left( x^q e^{-x^2} 
\right) \right|_{x=\frac{\sqrt{\tau}}{2}} \, .
\ee
Replacing in (\ref{high}), solving the $\tau$-integral 
and the remaining  $n$-sum, we get
\[
\left. s(p,q; \epsilon = 0) \right|_{z \gg 1} =
\]
\[
\frac{(-2 z)}{p-2} \frac{\partial}{\partial \mur}
\Re \left[
\left( \frac1{2 \pi z} \right)^{p-q-3}
\frac{\Gamma \left(\frac{p-q-3}{2}\right) 
\Gamma \left( \frac{q+1}2 \right)}{\Gamma \left( \frac{p-2}2 \right)}
\zeta \left( p-q-3 , \frac12 - i \frac{\mur}{2 \pi z} \right) \right. +
\]
\be
\left.
2 \sum_{\ell =0}^{\infty}
\frac{(-1)^\ell}{\ell !} \left(
\frac1{2 \pi z} \right)^{p-2+2 \ell} 
\frac{\Gamma \left( \frac{p-2 + 2\ell}{2}\right)}{\Gamma \left( 
\frac{p-2}2 \right)} \zeta \left( p-2+2 \ell, \frac12 -i 
\frac{\mur}{2 \pi z} \right) \zeta \left( - (q + 2 \ell), 
\frac12 \right) \right] \, ,
\label{gran-z}
\ee
for $p \ge 4 $.

\section{Numerical evaluations and applications}
\label{sec-aplicaciones}

The results obtained in the previous sections allow for the 
complete evaluation of the mean fermionic number 
and the $\mu-$dependent part of the Gibbs free 
energy for the fermionic field we are considering.

In order to obtain $N$, at fixed $z$, we can numerically 
evaluate $\tilde{N_1}$ from equations (\ref{eneuno}), 
(\ref{general}) and (\ref{desdeb}), and $\tilde{N_2}$ 
from (\ref{enedosdes}) and (\ref{limite}) to (\ref{ese2}). 

\begin{figure}
\epsffile{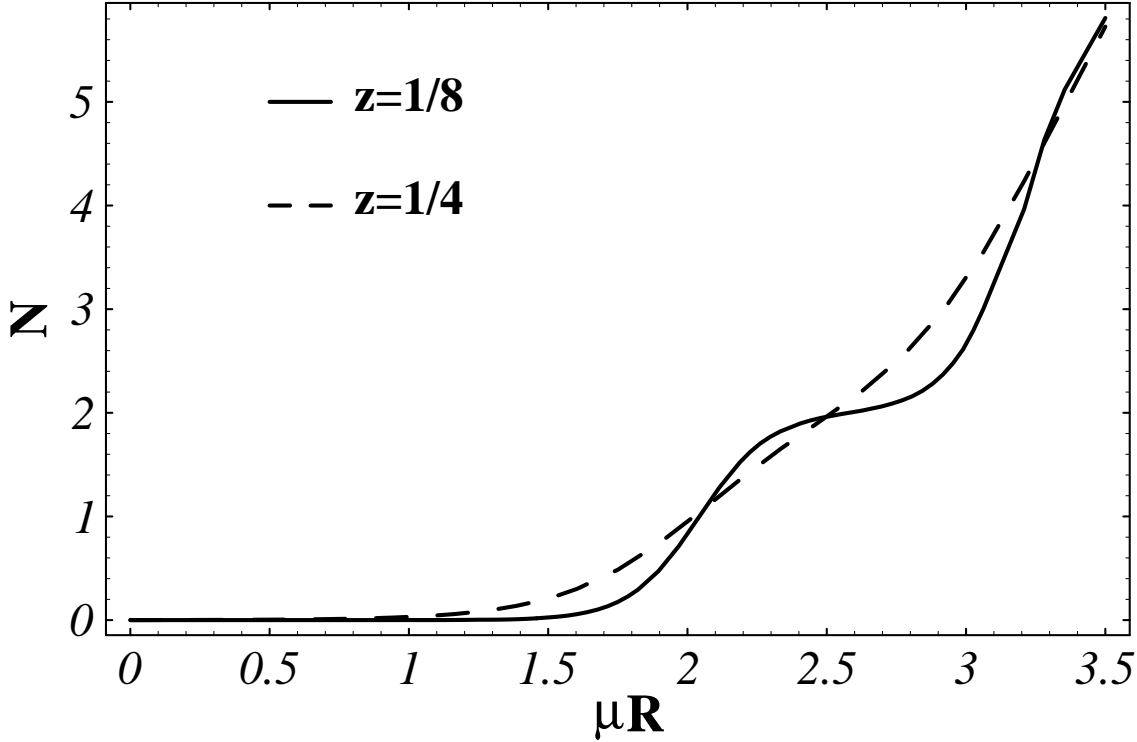}
\caption{Fermionic mean number for $z=1/8,1/4$ }
\label{fig-1}
\end{figure}

Adding these contributions to $N_0$, equation (\ref{ene0}), we obtain 
the fermionic mean number. The result is 
shown in Fig.~\ref{fig-1}, for $z=1/8, 1/4$

Notice that $N(z,\mur)$ has the expected behavior with $\mur$, since it shows 
steps at the adimensionalized eigenvalues of the Dirac Hamiltonian 
\cite{mulders}. 

\bigskip

Numerical evaluations become more difficult when $z$ grows up. But, in
the $z>1$ region, the asymptotic approximation for $\tilde{N_2}$ derived from
(\ref{gran-z}) can be used. Moreover, in this region $\left| \tilde{N_1}
\right| $ decreases with $z$. In fact, the first term of (\ref{gran-z})
suggests that $\tilde{N_1}$ contributes with a term of order $z^{-3}$. So, up
to this order, we obtain, 
\be N={{4\pi}\over{9}} \left(
\bar{\mu} z^2 + {{\bar{\mu}^3}\over{\pi^2}} \right) - \frac{2}{3 \pi} \mur
+ \frac{7 \zeta (3)}{126 \pi^3 z^2} \mur + O(z^{-3}),
\label{numero-gran-z} 
\ee

The first two terms in the right hand side of (\ref{numero-gran-z}),
proportional to $R^3$ and $R$ respectively, reproduce equation (27) in
Ref. \cite{mustafa}, where only the first finite size corrections to the
free energy are taken into account.  In the approximation leading to
(\ref{numero-gran-z}) we were able to get the next, $1/R$, term. 

\bigskip

Finally, the Gibbs free energy can be obtained by integrating the mean number,

\be
R \left( G (z,\mur) - F(z) \right) = 
- \int_0^{\mur} N(z,\mur \prime) \, d\mur \prime \, ,
\label{gibbs}
\ee
where $F(z)$ stands for the Hemholtz free energy of the system. In fact,
$F(z)$ has been studied in Ref. \cite{mitdef}, so the expression above
completely determines $G(z,\mur)$. 

\begin{figure}
\epsffile{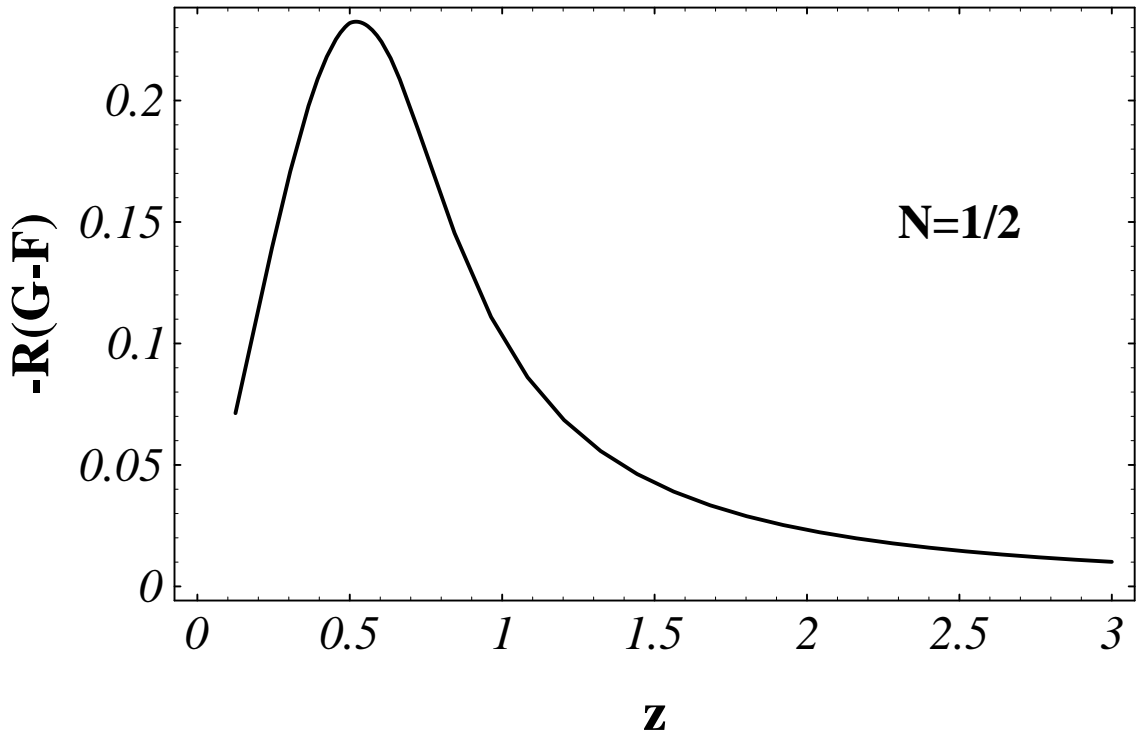}
\caption{Gibbs free energy at fixed fermionic number $N=1/2$}
\label{fig-2}
\end{figure}

If the fermionic number is fixed, in media, to $N$, one can obtain (numerically) the
chemical potencial as a function of $z$, $\mur (z,N)$. Replacing in
(\ref{gibbs}) we get
\be
R G (z,\mur (z,N))  =
-  \left[ \int_0^{\mur (z, N)}
d \mur \prime \, N (z, \mur \prime ) - R F(z) \right]
\label{dife}
\ee
The numerical evaluation of (\ref{dife}) for $N=1/2$, in a wide range of
values of $z$, is shown in Fig.~\ref{fig-2}. 

\subsubsection*{Application to the model of quark-gluon droplet}

We will now apply our results to the study a simple model of hadrons: a
static spherical droplet of confined quarks and gluons. 

Considering the degeneracies of flavor, $N_f =2$, and color, $N_c = 3$, we
obtain, from (\ref{dife})

\be
G_{\rm quarks} =
- \frac{N_f N_c}{R} \left[ \int_0^{\mur (z, N=3)}
d \mur \prime \, N (z, \mur \prime ) + z (- \beta F_{\rm fermions}(z)) \right]
\ee
where $- \beta F_{\rm fermions}(z)$ is given in section III of
Ref.~\cite{mitdef}. 

On the other hand, in the 1-loop approximation, the gluon field can be
described as $8$ copies ($N_g = 8$) of the Abelian gauge field studied in
Section IV of Ref.~\cite{mitdef}. So,

\be
G_{\rm gluons} = - \frac{N_g}{R} z \left( - \beta F_{\rm gauge} \right)
\ee 

Finally, a bag stabilizing term, $BV$, is introduced. The bag constant is
fixed to insure the pressure equilibrium when temperature reaches the
critical value (which we take as $T_0 = 150 {\rm MeV}$).  From the leading
volume terms, we obtain
\be
B= \frac{37}{90} \pi T_0^4 \, .
\ee

\begin{figure}
\epsffile{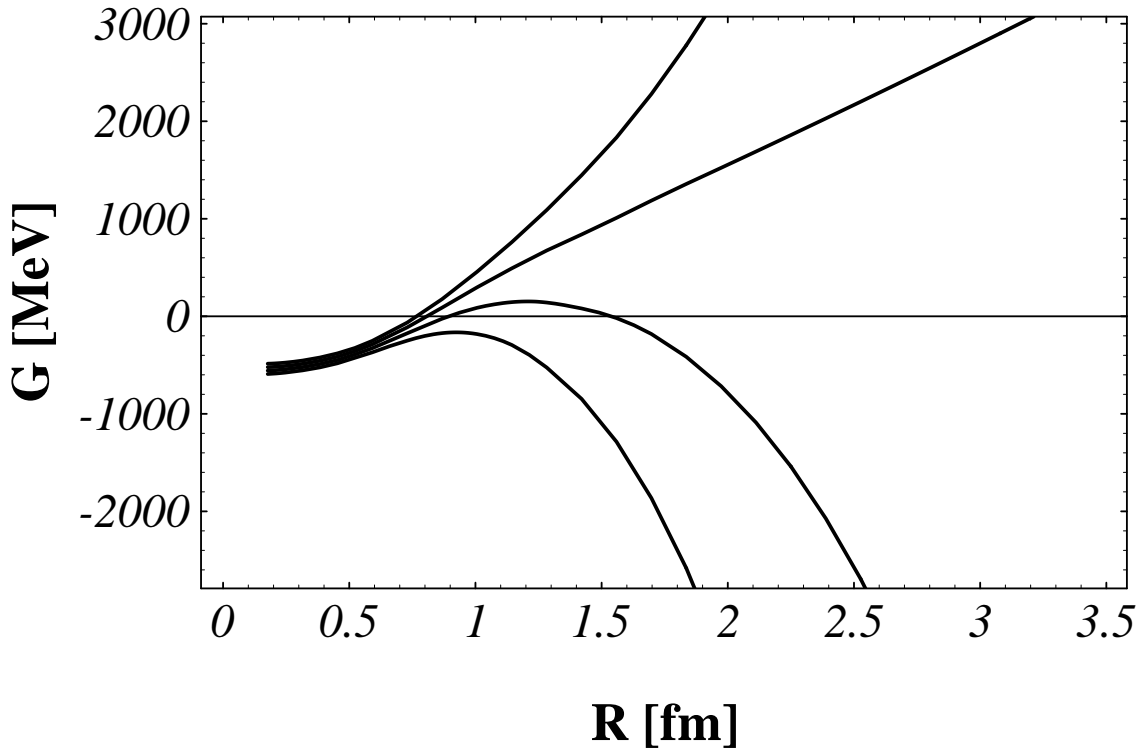}
\caption{Gibbs free energy for the quark-gluon droplet. From top to 
bottom: $T=140,150,160,170 {\rm MeV}$} 
\label{fig-3}
\end{figure}

When the Casimir energies of quarks and gluons (obtained in Ref.~\cite{mitdef} in the
framework of an analytic regularization), which can be written as $C/R$ with an
undetermined coefficient $C$, are not considered, the remaining
part of the total Gibbs free energy,
\be
G (z, \mur (z, N=3)) = G_{\rm quarks} + G_{\rm gluons} + BV
\ee
is as shown in Fig.~\ref{fig-3} for different values of the temperature.

The situation we find is consistent with the first order transition of a
hadronic bag to a deconfined quark-gluon phase described in
\cite{mustafa}: For each positive value of $C$, and for $0 \le T < T_0$, an 
absolute minimum of $G (z, \mur (z, N=3))$ appears at a
finite value of $R$, corresponding to a stable bag configuration. When
the temperature reaches $T_0$, this becomes only a local minimum 
representing a metastable state. When $T$ grows up, the local
minimum and the local maximum approach each other. Finally, there is a critical
temperature $T_c > T_0$ such that for $T>T_c$ there is no local
minima at finite $R$, and only the deconfined phase can exist. 

\bigskip

The computational techniques employed in this paper can, in principle, 
also be 
applied to more refined models of hadrons, such as the hybrid 
chiral bag 
model \cite{vento,rhoreport}. In Ref. \cite{cheshire} the behavior of the 
total energy at $T=0$ and $\mu=0$ has been analyzed, showing a remarkable 
independence on the bag radius, in agreement with the so called Cheshire 
Cat Principle \cite{rhoreport}. 
This justifies the proposal in \cite{deconf,loewe}, 
of looking for a deconfining transition only at the temperature (and $\mu$) 
dependent part of the free energy. 

The Helmholtz free energy ($\mu=0$) of a 
chiral bag can be obtained from the results in 
\cite{mitdef,bolsa-quiral}. 
The method developed in the present paper could be applied to study the
$\mu$ dependent part of the Gibbs free energy for this system. Notice 
that, for
this model, boundary conditions (which depend on the chiral angle) make the evaluation of the Green's
function more difficult, since there symmetry corresponds to the diagonal subgroup of
$SU(2)_{\rm rotational} \otimes SU(2)_{\rm isospin}$ (see 
Ref.~\cite{zahed} for the $T=0$ case). We will report on this subject elsewhere. 
\section*{Acknowledgements}

This work was supported in part by Fundaci\'on Andes -- Antorchas under Contract No. C-12777-9, CONICET (Argentina), and FONDECyT (Chile) under Grant No. 1950797. 


\end{document}